\documentclass[aps,prd,superscriptaddress,preprint]{revtex4}

\usepackage{float}
\usepackage{amsmath,amssymb}
\usepackage[dvips]{graphicx}
\DeclareGraphicsExtensions{.jpg,.mps,.pdf,.png,.eps} 
\usepackage{color}
\usepackage{hyperref}

\newcommand{\covu}[1]{\nabla^{#1}}
\newcommand{\covd}[1]{\nabla_{#1}}
\newcommand{\riemd}[4]{R_{#1#2#3#4}}
\newcommand{\riemu}[4]{R^{#1#2#3#4}}
\newcommand{\riccid}[2]{R_{#1#2}}
\newcommand{\ricciu}[2]{R^{#1#2}}

\newcommand{\scalricci}{\riccid{\alpha}{\beta}\ricciu{\alpha}{\beta}}
\newcommand{\scalriem}{\riemd{\alpha}{\beta}{\mu}{\nu}\riemu{\alpha}{\beta}{\mu}{\nu}}

\begin{document}

\title{Scale-invariant inflation with 1-loop quantum corrections}

\author{Silvia Vicentini}
\email{silvia.vicentini@unitn.it}
\author{Luciano Vanzo}
\email{luciano.vanzo@unitn.it}
\author{Massimiliano Rinaldi}
\email{massimiliano.rinaldi@unitn.it}

\affiliation{Department of Physics, University of Trento\\ Via Sommarive 14, 38123 Trento, Italy}
\affiliation{INFN - TIFPA \\ Via Sommarive 14, 38123 Trento, Italy }


\begin{abstract}
\noindent We study the quantum corrections to an inflationary model, which has the attractive feature of being classically scale-invariant. In this model, quadratic gravity plays along a scalar field in such a way that inflation begins near the unstable point of the effective potential and it ends at a stable fixed point, where the scale symmetry is broken and a fundamental mass scale naturally emerges. We compute the one loop corrections to the classical action on the curved background of the model and we report their effects on the classical dynamics with both analytical and numerical methods. 
\end{abstract}

\maketitle


\section{Introduction}

\noindent The recent Planck survey \cite{planck} provided a wealth  of observational data that allowed to put severe constraints on the space of inflationary models. Among these, the Starobinsky model \cite{star} results to be one of the most consistent with observations. This model is attractive because inflation is driven by a scale-invariant term $R^2$, in line with the fact that scalar and tensor perturbations are nearly scale-invariant. When inflation ends, the quadratic term becomes subdominant with respect to $M^2R$,   and so we are left with a Universe which has  a (Planck) mass scale $M$, in agreement with what we observe today \cite{defelice}. 

In this paper we investigate quantum corrections to a quadratic derivative model of inflation, which is presented in \cite{classic,tambalo}, in order to see whether its viability is preserved. This classical model is particularly attractive because it describes a scale-invariant inflationary phase, which ends in a scale-dependent fixed point of the action, as first explored by \cite{cooper}. This is particularly convenient for the same reasons that are given above for the Starobinsky model, although, here, the scale-dependence is achieved dynamically via spontaneous symmetry breaking. An analysis of the inflationary phase for the case of the Higgs field in place of a scalar field has been investigated in \cite{Gundhi:2018wyz}; in particular, a renormalization group driven quartic coupling has been considered. Other relevant contributions can be found in \cite{Kurkov:2013gma,Myrzakulov:2016tsz,inf3,inf1,inf2,lalak} and references therein, focusing on various aspects of the quantum corrections to inflation, like quantum anomalies, the influence of gauge fields, dark matter or $f(R)$.  Scale-invariant gravity in $f(R)$ was investigated also in the context of classical black holes \cite{bh1,bh2}.

Quantum corrections are known to break the conformal symmetry \cite{shapiro,parker} and, in particular, scale symmetry, since a regulation scheme necessarily introduces mass scales in the action.  A detailed study showing the effect of gauge degrees of freedom in forming a symmetry breaking scalar condensate has been recently presented in \cite{Kubo:2018kho}. To make the model described in \cite{classic} more robust, we need to check  that the one-loop contributions are suppressed with respect to the classical action and so the breaking of scale symmetry is mild, at least on-shell, during inflation.  We should note that conflicting claims about a conformal anomaly exist, since scale-invariant regulators have been used in recent articles, e.g. the new approach presented in \cite{scaleinv2}, and the field-dependent mass scale in \cite{scaleinv3} or \cite{scaleinv1}. Calculations are carried out on curved spacetime, in which gravity is kept classical (including the $R^{2}$ term) and all other fields are quantized: this theory has proven to be very effective in predicting physical phenomena such as the Hawking radiation and the formation of large-scale structures in the Universe \cite{birrell,parker}. However, as we will see, the presence of tachyonic instabilities in the conditions required for inflation may  actually restrict the validity of the method. 

We end this introduction with few considerations on the origin of the additional scalar field in the action \eqref{lagrangian}. We think that it can be motivated more strongly from low energy particle physics, rather than inflation, because the Standard Model Lagrangian is exactly scale invariant were it not for the Higgs mass, an old remark probably due to W. Bardeen \cite{WB95}. This suggested to him the idea that the Higgs mass could emerge via broken scale invariance due to the vacuum expectation value of one or more scalar fields, in such a way that the smallness of the Higgs mass would be technically natural. The preferred scalar field is not necessarily the Higgs itself. Another hint comes from the nearly scale invariant spectrum of primordial fluctuations. However, the most natural framework where the appearance of a low energy scalar field is actually predicted is string theory, with its dilaton field.  The low energy effective action of string theory actually contains all the terms (plus many more) of the action \eqref{lagrangian}, in what it is called the string metric by string theorists, and the Jordan frame metric by cosmologists.  The Brans-Dicke theory can also be considered as another instance of the dilaton field, but with a different coupling to the metric.  The dilaton field is always part of the string low energy action and as a consequence there is, strictly speaking, no equivalence principle and thus no way to single out a preferred metric. This is one motivation to include the scalar field as the conformal part of a new metric, known to everybody as the Einstein frame metric. The stringy stuff accompanying the dilaton is omitted in this paper, on the grounds that inflation generically ends so fast that all interactions, except the gravity-scalar sector, are frozen because the corresponding rates are much more slower than the expansion rate (the Gamow argument).  

This paper is organized as follows. In Sec.\ \ref{classic} we give a brief introduction to the classical model presented in \cite{classic}, in order to highlight the principal results, which are to be compared with the quantum ones.  Then, in Sec.\ \ref{1loop} we compute the one loop correction along with renormalization group equations. We numerically study the dynamics and use a method presented in \cite{herr}, which allows to find  approximate quantitative results. As long as we need qualitative bounds, this approximation suffices. To go beyond this approximation, we will use numerical methods. We finally discuss our findings and future work in \ref{conc}. Most cumbersome formulae are contained in the appendix.


\section{The classical model}\label{classic}

\noindent We consider the model presented in \cite{classic} with action
\begin{equation}
\label{lagrangian}
S=\int d^4x\sqrt{-g}\left[\frac{\alpha}{36}R^2+\frac{\xi \phi^2 R}{6}-\frac{1}{2}(\partial \phi)^2-\frac{\lambda}{4}\phi^4\right]\,,
\end{equation}
where $\xi>0$, $\lambda>0$. This action is scale-invariant, i.e. invariant under
the transformations
\begin{equation}\bar{g}_{\mu\nu}(x)=g_{\mu\nu}(\ell x)\,,\qquad \bar{\phi}(x)=\ell\phi(\ell x)\,.\end{equation} It is also invariant under the internal Weyl symmetry 
\begin{equation}\bar{g}_{\mu\nu}(x)=L^2g_{\mu\nu}(x)\,,\qquad \bar{\phi}(x)=L^{-1}\phi(x)\,.\end{equation} 
From now on we choose, as background metric, a flat Robertson-Walker line element with signature $(-,+,+,+)$. The three parameters $\alpha,\xi,\lambda$ are dimensionless free parameters.

The effective classical potential 
\begin{equation}\label{effpot}
V(\phi)=-\frac{\xi \phi^2 R}{6}+\frac{\lambda}{4}\phi^4
\end{equation}
has two stationary points at $\phi=0$ and $\phi=\pm 2\sqrt{\frac{\xi}{\lambda}}H_1$ for some constant $H_{1}$. From the equations of motion (which are of second order in $\phi$ and $H$), we find that the stationary points are also fixed points of the dynamical system in the phase space $(\phi,H)$. In particular, it turns out that the first is a saddle point and the second is a stable attractor. When the point in phase space reaches the stable fixed point scale-symmetry spontaneously breaks, in the sense that the scalar field settles at a non-vanishing value. If we further impose the constraint $\alpha=\xi^{2}/\lambda$ the quadratic curvature term exactly cancels the quartic potential at the stable fixed point. Here, \eqref{lagrangian} reduces to the usual Einstein-Hilbert action with a mass scale determined by the value of scalar field at the minimum of the potential. This mass can naturally be identified with the Planck mass.

We can solve the linearized system of equations near the fixed points in terms of the number of e-foldings $N=\log a$. Close to the saddle point we find
\begin{equation}
\label{fp1}
H(N)=c_1+c_2e^{-3N}\,,
\end{equation}
\begin{equation}
\label{fp2}
\phi(N)=c_3e^{\left(-\frac{3}{2}+\frac{1}{2}\sqrt{9+16\xi}\right)N}+c_4e^{\left(-\frac{3}{2}-\frac{1}{2}\sqrt{9+16\xi}\right)N}\,.
\end{equation}
Close to the stable attractor instead we have
\begin{equation}
\label{fp3}
H(N)=c_1+c_2e^{-3N}+e^{-3/2N}(c_3S(N)+c_4C(N))\,,
\end{equation}
\begin{equation}
\label{fp4}
\phi(N)=\frac{\xi}{\lambda}\left[2c_1+\frac{c_2}{2}e^{-3N}+\frac{\xi}{2(1+2\xi)}e^{-3/2N}((2Kc_4-5c_3)S(N)-(5c_4+2Kc_3)C(N))\right]\,,
\end{equation}
where $K=\frac{1}{2}\sqrt{7+64\xi}$ and $S(N)=\sin(KN)$, $C(N)=\cos(KN)$.



With these approximations it becomes clear that, in the proximity of the saddle point, the evolution of the Universe is quasi-de Sitter and one finds the following relation between the number of e-foldings required by inflation ($\Delta N$) and the initial values for the dimensionless ratio $\frac{H_{i}}{\phi_{i}}$:
\begin{equation}
\label{deltan}
\Delta N=\frac{1}{2}\ln\left[\frac{(2\xi-3)H_i^2}{\lambda\phi_i^2}\right]\implies \frac{H_i}{\phi_i}\simeq \exp(\Delta N-9).
\end{equation}
The latter relation is obtained, in particular, if we assume ``physical'' values of the couplings $\xi=1$  and $\lambda=10^{-8}$ \footnote{We consider $\xi=1$ physically plausible since $\xi=1/2$ is the conformal coupling for a ghost scalar field. On the other hand, $\lambda=10^{-8}$ provides a viable initial condition for the evolution of the Universe at the end of the inflationary phase. }. With these values, we ensure that, when the system settles at the stable fixed point, $\frac{\xi}{3}\phi^2=M_p^2=(8\pi G)^{-1}$.  The observational constraint $\Delta N\geq 60$, needed to solve the flatness and horizon problems \cite{guth}, is satisfied if inflation begins at a point in the phase space close enough to the unstable fixed point. Moreover, numerical computations show that, after inflation ends, the system settles in the stable fixed point in few e-foldings, during which both $H$ and $\phi$ undergo damped oscillations, able to give rise to reheating (see \cite{classic} for details).

To obtain an approximate value for the inflationary spectral indices, we transform the action into the more familiar Einstein frame. Let us consider the Lagrangian
 \begin{equation}
 \label{jordan}
 \mathcal{L}=\chi R-\frac{(\partial \phi)^2}{2}-\frac{\alpha \varphi^2}{36}-\frac{\lambda}{4}\phi^4=\chi R-\frac{(\partial \phi)^2}{2}-\frac{\lambda}{2}\phi^4+
 3\frac{\xi\chi\phi^2}{\alpha}-\frac{9}{\alpha}\chi^2\,,
 \end{equation}
 where \begin{equation}\chi=\frac{\alpha}{18}\varphi+\frac{\xi}{6}\phi^2\,.\end{equation}
 This is the same as Eq. \eqref{lagrangian}, since the equation of motion for $\varphi$ gives $\varphi=R$. We reparametrize the fields with a conformal transformation $\bar{g}_{\mu\nu}=\frac{2}{\mathcal{M}^2}\frac{\partial \mathcal{L}}{\partial R}g_{\mu\nu}$ \footnote{A conformal transformation in a model with dynamical gravity amounts to a redefinition of the fields, as thoroughly explained in \cite{herzberg}.}. All other tensors transform accordingly (as given in \cite{dabrovski}). One then finds the Einstein frame action
\begin{equation}
\label{einst}
S=\int d^4x \sqrt{|\bar{g}|}\left[\frac{\mathcal{M}^2}{2}\bar{R}-\frac{(\partial_{\mu}\psi)^2}{2}-\frac{(\partial_{\mu}\phi)^2}{2}\exp\left(-\frac{\sqrt{2}\psi}{\sqrt{3}\mathcal{M}}\right)-W(\phi,\varphi)\right]\,,
\end{equation}
where the potential is given by
$$W(\phi,\varphi)=\frac{9\lambda \mathcal{M}^4}{4\xi^2}+\frac{\lambda \phi^4}{2}\exp\left(-\frac{2\sqrt{2}\psi}{\sqrt{3}\mathcal{M}}\right)-\frac{3\lambda \mathcal{M}^2\phi^2}{2\xi}\exp\left(-\frac{\sqrt{2}\psi}{\sqrt{3}\mathcal{M}}\right)\,,$$
and where we redefined the ``scalaron'' field $\psi\equiv \sqrt{6}\mathcal{M}\ln\Omega$. Note that the mass parameter $\mathcal{M}$ is completely arbitrary and, although it is not apparent, scale-invariance can be preserved in the Einstein frame if $\mathcal{M}\rightarrow L^{-1}\mathcal{M}$ under scale transformations. The Einstein frame action \eqref{einst} describes the dynamics of two scalar fields besides the Einstein term, this can be reduced to single field inflation, as in \cite{Karam:2018mft} and \cite{tambalo}.

As in the Jordan frame, there are two stationary points. Interestingly, they satisfy a universal scaling between the Hubble functions calculated at the two fixed points given by  $\frac{\bar H_{\rm unst}}{\bar H_{\rm st}}=\sqrt{2}$. The slow roll parameters are
\begin{equation}\epsilon=-\frac{d\bar{H}/d\bar{t}}{\bar{H}^2}\sim\frac{\mathcal{M}^2}{2}\left(\frac{\partial W}{\partial \psi}\frac{1}{W}\right)^2 \,,\end{equation}
\begin{equation}\eta=\frac{d^2\psi/dt^2}{\bar{H}d\psi/dt}\sim\epsilon-\frac{\mathcal{M}^2}{W}\frac{\partial^2W}{\partial \psi^2}\,,\end{equation}
and the number of e-foldings is
\begin{equation}
\label{efolds}
\bar{N}=\int \bar{H}d\bar{t}\sim-\frac{1}{\mathcal{M}^2}\int d\psi\,W\left(\frac{\partial W}{\partial \psi}\right)^{-1}.
\end{equation} 
Since inflation occurs near the unstable fixed point we can expand the potential for $\frac{\phi}{\mathcal{M}}\ll1$ and we find
\begin{equation}
\label{result}
\epsilon\sim\frac{3}{4N^2}\,,\qquad \eta\sim\epsilon+\frac{1}{N}\implies n_s=1-2\eta-4\epsilon\sim 1-\frac{2}{N}+O\left(\frac{1}{N^2}\right)\,,
\end{equation}as in the Starobinsky model \cite{defelice}. For more detailed results, see \cite{tambalo} and also, for a more comprehensive class of models, \cite{canko}.  

In this brief summary we have described an inflationary model where scale invariance is broken dynamically and classically. The spectral indices are very similar to the ones predicted by the Starobinsky model but with a different scale invariance breaking mechanism. We now turn to the quantum corrections that are expected to arise in the scalar field sector. 


\section{One loop effective action}\label{1loop}

\noindent We now compute quantum corrections to the classical model and we choose the Jordan frame, where calculations are simpler. This choice implicitly amounts to consider the ``scalaron'' degree of freedom as classical. Thus,  calculations are carried out in the framework of semi-classical gravity, which is introduced in \cite{birrell} and \cite{parker} (see also the comprehensive DeWitt's book \cite{DeWitt:2003pm}).

 The usual approach to divergences that appear in an effective action computed on curved spacetime is to treat  them with dimensional regularization, which is known to break scale symmetry, as any other commonly-used regulator.
 
We consider the action \begin{equation}
\Gamma[g_{\mu\nu},\phi]=\Gamma^{[0]}[g_{\mu\nu},\phi]+\Gamma^{[1]}[g_{\mu\nu},\phi],
\end{equation}
where
\begin{equation}
\Gamma^{[0]}[g_{\mu\nu},\phi]=S_m[g_{\mu\nu},\phi]+S_g[g_{\mu\nu},\phi]+\delta S[g_{\mu\nu},\phi]
\end{equation}
is the tree level action ($S_{g}+S_{m}$) plus the counterterm action ($\delta S$). The term
\begin{equation}\label{gamma1}
\Gamma^{[1]}[g_{\mu\nu},\phi]=-\frac{i}{2}\hbar\,\text{Tr}\ln( - G(x,x'))
\end{equation}
is the one loop correction, which depends on the biscalar propagator $G(x,x')$ in the background fields $(g_{\mu\nu},\phi)$ \footnote{From now on, we set $\hbar=1$.}.
The propagator is expanded in Riemann normal coordinates  and \eqref{gamma1} can be evaluated with standard techniques to give
\begin{equation}
\label{regularized}
\Gamma^{[1]}[g_{\mu\nu},\phi]=\int d^nx|g(x)|^{1/2}\,\frac{1}{2(4\pi)^{n/2}}\left(\frac{M^2}{\mu^2}\right)^{(n-4)/2}\sum_{j=0}^{+\infty}a_j(x,x)(M^2)^{n/2-j}\Gamma(j-n/2)\end{equation}
in $n$ spacetime dimensions. Here, $\mu$ is the external mass scale that appears due to dimensional regularization and $M^2$ is the  mass associated to the quantum perturbation, determined below. As stated in \cite{parker,DeWitt:2003pm}, $M^2$ should have the Feynman prescription $M^2-i\epsilon$ as long as it is positive, to make the integral representation of the semi-classical propagator expansion converge. Equation \eqref{regularized} is valid when spacetime is slowly expanding, meaning that 
\begin{equation}
\label{adiabaticcondition}
\frac{(\dot{a}/a)^2}{M^2}\ll 1\,, \qquad \frac{\ddot{a}/a}{M^2}\ll 1\,,
\end{equation} 
where $a(t)$ is the scale factor in a flat Robertson-Walker metric. 

The propagator associated to quantum fluctuations satisfies
\begin{equation}
\left(-\Box +3\lambda\phi^2-\frac{\xi}{3}R\right)G(x,x')=\frac{\delta(x-x')}{\sqrt{-g}}\,.
\end{equation}
and, since the scalar field is massless,  we set
\begin{equation}
\label{mass}
M^2=3\lambda\phi^2 -\frac{(2\xi+1)}{6}R\,,
\end{equation}
so that the quantum fluctuation satisfy
\begin{equation}\left(-\Box+M^2+\frac{R}{6}\right)\varphi=0\,.\end{equation}
The $\frac{R}{6}$ term in \eqref{mass}  is introduced because it allows to sum part of the Riemann expansion of the propagator when given in terms of $M^2+\frac{R}{6}$ ($R$-summed propagator \cite{parker}).
Actually, the classical evolution implies that $M^2$ changes from being negative  to positive when going from the unstable point to the stable one.

It is unclear to us how to proceed in this case. Going back in time, the scalar effective mass $M^2$ becomes negative  when $\phi^2$ drops below a quantity of order $R/\lambda\sim H^2/\lambda$ (with $\xi\sim 1$), which may happen  to be close to the unstable fixed point, and the field disturbances become tachyonic during few e-foldings of order $2N\sim\log (H^2/\lambda\phi_{0}^2)$, where $\phi_{0}$ is the initial value of the field. So we have to keep this number well below that required for inflation ($N\geq 50$). Comparing with Eq.~\eqref{deltan}, this is barely satisfied in the growing field regime. As a qualification, the term tachyonic has nothing to do here with superluminal propagation. Rather, it refers to an instability, since tachyonic fields have unbounded energy from either side. Indeed, it is exactly this tachyonic instability that makes the field to grow exponentially till the minimum of the potential. In the crossover region $M^2\sim 0$, the disturbances are nearly massless. Classically, a tachyonic field is not seriously problematic. Quantum mechanically things are very different and indeed it has been argued that this provides a very efficient preheating process \cite{Felder:2000hj,Felder:2001kt}.  However, it is true that we have here a variable mass and the effects of the $R^2$ term, that make all the difference. 

Unless otherwise stated, it is understood in the following that our analysis applies only outside the region of tachyonic instability in the field space, and consequently we will judge the results from the consistency requirement that the one-loop expansion really can be applied. The quantum treatment of the tachyonic regime is nevertheless very interesting  and  deserves further work. It is known that these theories are not unitary unless due care is taken of the interactions, and even quantization of the free tachyon field may result in a violation of Einstein causality over macroscopic scales. That said, we may come now to the heat kernel expansion.

The coefficients of the expansion in Eq. \eqref{regularized} up to second order are taken from \cite{mark} and are reported in the Appendix \ref{appA}. Up to second order we find
\begin{equation}
\begin{split}
\Gamma[g_{\mu\nu},\phi]=&\Gamma^{[0]}[g_{\mu\nu},\phi]+\int d^nx \frac{\sqrt{-g}}{64 \pi^2}\left\{-M^4\left[\log\left(\frac{M^2}{\mu^2}\right)-\frac{3}{2}+\frac{1}{n-4}\right]+\right.\\&\left.\qquad-2a_2(x,x)\left[\log\left(\frac{M^2}{\mu^2}\right)+\frac{1}{n-4}\right]\right\}\,,
\end{split}
\end{equation}
where the Euler-Mascheroni constant has been absorbed into $\mu^2$. Divergences are canceled by counterterms in $\delta S$ in the $\overline{MS}$-scheme, giving
\begin{equation}
\begin{split}
\label{onelooplagrangian}
\mathcal{L}[g_{\mu\nu},\phi]&=\frac{\alpha}{36}R^2+\frac{\xi}{6}\phi^2 R-\frac{(\partial_{\mu}\phi)^2}{2}-\frac{\lambda}{4}\phi^4+\epsilon_1 \riccid{\alpha}{\beta}\ricciu{\alpha}{\beta}+\epsilon_2\riemd{\alpha}{\beta}{\mu}{\nu}\riemu{\alpha}{\beta}{\mu}{\nu}+\\&+\frac{1}{64\pi^2}\left\{-M^4\left[\log\left(\frac{M^2}{\mu^2}\right)-\frac{3}{2}\right]+\frac{\riccid{\mu}{\nu}\ricciu{\mu}{\nu}-\riemd{\alpha}{\beta}{\mu}{\nu}\riemu{\alpha}{\beta}{\mu}{\nu}}{90}\left[\log\left(\frac{M^2}{\mu^2}\right)\right]\right\}.
\end{split}
\end{equation}
We see that we should add two new couplings $\epsilon_1$, $\epsilon_2$ at tree-level to account for all divergences and that the Lagrangian is well defined for positive $M^2$, picking an imaginary part for negative $M^2$ if the one-loop result could be trusted. 

In passing, we note that the sign of the imaginary part is determined if we remember that $M^2$ really is $M^2-i0$, so we must approach the cut of the logarithm from below, which would give the imaginary part of the effective action
\begin{equation}
{\Im}(\Gamma)=\int\frac{M^4}{64\pi}d\mu-\int\frac{\riccid{\mu}{\nu}\ricciu{\mu}{\nu}-\riemd{\alpha}{\beta}{\mu}{\nu}\riemu{\alpha}{\beta}{\mu}{\nu}}{5760\pi}d\mu\,, \quad d\mu=\sqrt{-g}d^nx\,,
\end{equation}
and $\exp[-\Im(\Gamma)]$ indicates the quantum instability we mentioned before. The first term cannot be trusted though. The curvature term possibly can, but should be supplemented with the metric curvature fluctuations due to the $R^2$ term (it is negative anyway in a FRW background). General results for curvature perturbations in modified gravity can be found in  \cite{Ruf:2017bqx} and for a constant curvature background in \cite{Bamba:2014mua}. We emphasize that the divergent part must always be real, and that the imaginary part is finite. In this regard, one should remember that Eq.\ \eqref{onelooplagrangian} does not provide a faithful description when conditions \eqref{adiabaticcondition} are not met, which surely happens when $M^2\sim 0$.  This means that Eq.\ \eqref{regularized} is not valid when $M^2$ is vanishing, but we can expect its real part to have a smooth massless limit \cite{parker} which would be valid in the crossover region. 

The modified equations of motion thus are 
\begin{equation}
\label{eom1}
\alpha H^2(2HH''+H'^2+6HH')+2\xi H^2\phi\phi'-\frac{\phi'^2}{2}H^2+\frac{\phi^4}{4}(4\xi H^2-\lambda\phi^2)+\text{Q}_1=0\,,
\end{equation}and
\begin{equation}
\label{eom2}
H^2\phi''+(HH'+3H^2)\phi'-2\xi\phi HH'-\phi(4\xi H^2-\lambda\phi^2)+\text{Q}_2=0\,,
\end{equation}
where $Q_{1}$ and $Q_{2}$ contain all the quantum corrections  and are explicitly given in the Appendix \ref{appB}.

Since the effective action should be independent on the mass scale $\mu$, we have $\mu\frac{d\Gamma}{d\mu}=0$, from which retrieve the energy dependence of the renormalized couplings. These are expressed in terms of the beta functions $\beta_{q_i}\equiv\mu\frac{dq_i}{d\mu}$, where $q_i$ is a generic coupling constant. We then find
\begin{equation}
\label{runcoupling}
\beta_{\lambda}=\frac{9\lambda^2}{8 \pi^2}\,,\qquad\beta_{\xi}=\frac{3\lambda(2\xi+1)}{16 \pi^2}\,,\qquad\beta_{\alpha}=-\frac{(2\xi+1)^2}{32\pi^2}\,,\end{equation}
$$\beta_{\epsilon_1}=\frac{1}{2880\pi^2}\,,\qquad\beta_{\epsilon_2}=-\frac{1}{2880\pi^2}\,.
$$
As a check, the first beta function matches exactly the standard beta function of the quartic interaction, and does not feel the curvature in this approximation. 
Taking the classical reference value $\lambda_0=10^{-8}$ and solving Eq.\  \eqref{runcoupling} for the couplings (see Appendix \ref{appC}), we see that the running  of $\xi(\mu)$ and $\alpha(\mu)$ is suppressed by $\lambda_0$. Moreover, a factor $\frac{1}{2880\pi^2}$ appears in the running of $\epsilon_1(\mu)$ and $\epsilon_2(\mu)$, which are then also suppressed for sufficiently small values of $\mu$. $\beta_{\epsilon_{1,2}}$ are the residues of the poles in the one loop effective action, as predicted by \cite{shapiro}. Moreover, we recover asymptotic freedom conditions, in the infrared for $\lambda$, $\epsilon_{1}$ and $\xi$, in the ultraviolet for $\alpha$ and $\epsilon_{2}$: as energy grows ($\mu\rightarrow +\infty$), the self coupling $\lambda$ runs toward a Landau pole, $\xi$ flows to its conformal value $\xi=-1/2$ and the gravitational couplings $\alpha$ and $\epsilon_2$ become weaker and weaker. Due to different runnings, the constraint $\alpha=\xi^2/\lambda$ no longer holds. The divergences for $\mu\rightarrow 0$ of $\alpha(\mu)$, $\epsilon_1(\mu)$ and $\epsilon_2(\mu)$ reflect the infrared divergences that typically appear in a massless theory.

\subsection{Numerical solution}

\noindent We choose the external mass scale as  \begin{equation}
\label{extmass}
\mu^2=M^2(\phi,\lambda(\mu),\xi(\mu),R)\,,\end{equation}  so it is time-dependent. This choice is very convenient since it makes all logarithms vanish, but it also makes the renormalization group Eq.\ \eqref{runcoupling} time-dependent and so it has to be solved along with the equations of motion. Moreover, it makes the whole system of equations implicit, making quantitative predictions very difficult \footnote{These will be given in the pseudo-optimal energy scale choice, see below.}. We choose, as initial conditions near the unstable de Sitter phase, the classical initial values $(H_0,\phi_{0})$ and the classical values for the couplings. The new couplings $\epsilon_1$ and $\epsilon_2$ are taken to be zero at the reference scale.

The numerical solution of the system shows that the dynamical evolution is very similar to the classical one, as shown in  fig.\ \ref{figure}. In particular, fields in the second de Sitter phase vary at most $\sim 1\%$ with respect to the classical case and couplings stay around their reference value. The mass scale changes little, staying in the range $0.4  \mu_0\leq \mu\leq1.5 \mu_0$  when the adiabatic approximation is valid. In this range, we have that the derivatives of a generic coupling $q_i$ in the number of e-foldings  satisfy the constraint \begin{equation}\bigg\lvert\frac{q_i'}{q_i}\bigg\rvert\lesssim 0.05\,.\end{equation} for each coupling $q_i$.

We now consider the same energy scale \eqref{extmass} but we set it in the Lagrangian \emph{before} taking the variation of the action to derive the equations of motion. This is the pseudo-optimal energy scale choice introduced in \cite{herr}, and it is an approximation since it gives $\frac{d\Gamma}{d\mu}\neq 0$. We find that the pseudo-optimal energy scale choice is a good approximation (as can be seen in Fig.\eqref{figure}) and that fields in the second de Sitter phase vary at most $\sim 5 \%$ with respect to the classical case.

To test the validity of the approximation introduced by the pseudo-optimal energy scale choice, the authors of  \cite{herr} propose to verify that  $\mu\frac{d\Gamma}{d\mu}\ll1$, but actually we find that   $\mu\frac{d\Gamma}{d\mu}\gg1$ for most of the evolution, as can be seen in Fig.\eqref{figure}.

We verified numerically when the adiabatic condition \eqref{adiabaticcondition} is met during the inflationary phase: as can be seen from Fig.\eqref{figure}, if we take $0.05$ as critical value for the adiabatic condition Eq.\eqref{adiabaticcondition}, we see that it is violated when the mass scale goes to zero ($N\sim 2$ in Fig.\ref{figure}), and mildly once after ($N\sim 2.3$). In this regime we can not make any prediction since the form of the effective action itself depends on this approximation. Whenever computing observables in the following, we ensure that they do not fall in these two lapses of time. Out of the adiabatic regime particles are created by the changing spacetime: these particles will decay in Standard Model particles in the oscillations around the stable fixed point. In principle they have a backreaction on the geometry \cite{parker}, but this is neglected here.


\subsection{Fixed points}

In the following we use the pseudo-optimal energy scale choice in order to compute some quantitative results.   We can also neglect the running of the couplings, as seen in the last paragraph. We find that the two fixed points are still present and are given by
\begin{equation}
 (H_0,0)\,,\qquad \left(H_1,\pm\sqrt{\frac{12\left(\frac{\xi }{3}-\frac{3}{64\pi^2}\lambda(2\xi+1)\right)}{\left(\lambda-\frac{27}{32 \pi^2}\lambda^2\right)}}H_1\right)\,.
 \end{equation}
These analytic expressions  match up to 3\% with the ones computed numerically in the full one-loop case. Moreover, they are a saddle (unstable) fixed point and a minimum, respectively. Near the unstable point we have
\begin{equation}H(N)=c_1+c_2e^{-B/AN}\,,\end{equation}
where \begin{equation}B=3+48\epsilon_1+84\epsilon_2+\frac{9}{128\pi^2}(2\xi+1)^2\,,\end{equation}
\begin{equation}A=1+12 (\epsilon_1+\epsilon_2)+\frac{3}{128\pi^2}(2\xi+1)^2\,.\end{equation}
The solution is close to the classical one due to the weak energy dependence of the couplings. Anyway the stability of the fixed points is preserved for arbitrary real values of $\xi$ as long as $\epsilon_1$ and $\epsilon_2$ are non-negative. $\phi(N)$ is as in Eq.\eqref{fp2}, but with
\begin{equation}\sqrt{9+16\xi}\rightarrow \sqrt{9+\left(16\xi-\frac{9}{4\pi^2}\lambda (2\xi+1) \right)}\,.\end{equation}
Growing and decaying modes could be spoiled by sufficiently large values of $\lambda$. Oscillatory modes appear when
\begin{equation}\xi<\frac{\frac{9}{4\pi^2}\lambda-9}{16-\frac{9}{2\pi^2}\lambda}\,.\end{equation} This is never verified for $\xi>0$ and $\lambda \ll 1$. In particular, taking $\xi>0$, oscillatory modes may appear when \begin{equation}\frac{-9+\frac{9}{4\pi^2}\lambda}{16-\frac{9}{2\pi^2}\lambda}>0\,,\end{equation} that is when $\lambda\in (32\pi^2,36\pi^2)$. For $\lambda=0.1$, which is used in the numerical solution, we obtain $\xi\leq-0.5$ as a critical value, so this point has the same stability of the "physical" one. Concerning the stable fixed point, we linearized and diagonalized numerically the system of differential equations, finding that only small deviations appear with respect to the classical result.


\subsection{Inflation}

\noindent We can compute the dependence of the number of e-foldings on the value of the fields in the unstable fixed point, as in Eq.\ \eqref{deltan}. We consider
\begin{equation}\epsilon_1=\frac{-H'}{H}=\frac{H^2\phi''+3H^2\phi'+\lambda\phi^3-4\xi\phi H^2-\frac{3}{128 \pi^2}(36 \lambda^2 \phi^3-24\lambda(2\xi+1)H^2 \phi)}{H^2(\phi'-2\xi\phi-\frac{36}{128\pi^2}\lambda(2\xi+1))}\,,\end{equation}
and, by imposing $\epsilon_1=1$ at $N_e$ (end of inflation) and by using Eqs.\ \eqref{fp1}, Eq.\eqref{fp2} for $N_e-N_i$, we find
\begin{equation}N_e-N_i\sim\frac{1}{2}\ln\left(\frac{H^2(-3-\frac{108\lambda}{128\pi^2}(2\xi+1)+2\xi)}{(\lambda-\frac{108}{128\pi^2}\lambda^2)\phi^2}\right)\,,\end{equation}
so the number of e-foldings for inflation has the same dependence on $\frac{H}{\phi}$ as in the classical case and, in a sufficiently small neighborhood of the unstable point, the condition $\Delta N\geq 60$ can always be met. Thus, also with quantum corrections, inflation can last long enough to satisfy the observational constraints.

Finally, we  have numerically verified  that the deviations from a null cosmological constant (with the constraint  $\alpha=\frac{\xi^2}{\lambda}$) around the stable fixed point are small, since they are just 4\% the value of the cosmological constant in the unstable de Sitter one.

\subsection{Spectral indices}

Regarding the computation of the spectral indices, the easiest way  is to rely on the same method that has been used in the previous section. The problem, however, is that the correspondence among Einstein and Jordan frame is not completely assessed at quantum level. Nevertheless, it can be argued that the two descriptions should match on-shell in order to have the correct S-matrix elements, see e.g.\  \cite{quantum1,Ruf:2017xon,quantum2} (on the equivalence of the two frames in the space of solutions see however \cite{maxeq}). With the pseudo-optimal energy scale choice the Lagrangian in the Einstein frame is
\begin{equation}
\label{einstein}
\begin{split}
S=&\int d^4x \sqrt{|g|}\left[\frac{\mathcal{M}^2}{2}\bar{R}-\frac{(\partial_{\mu}\psi)^2}{2}-\frac{(\partial_{\mu}\phi)^2}{2}\exp\left(-\frac{\sqrt{2}\psi}{\sqrt{3}\mathcal{M}}\right)-\frac{9 \mathcal{M}^4}{4\left(\alpha+\frac{3}{128\pi^2}(2\xi+1)^2\right)}\right.\\&\left.-\frac{\left(\lambda-\frac{3}{32\pi^2}\lambda^2\right) \phi^4}{2}\exp\left(-\frac{2\sqrt{2}\psi}{\sqrt{3}\mathcal{M}}\right)-\frac{3\left(\xi-\frac{3}{32\pi^2}\lambda (2\xi+1)\right) \mathcal{M}^2\phi^2}{2\left(\alpha+\frac{3}{128\pi^2}(2\xi+1)^2\right)}\exp\left(-\frac{\sqrt{2}\psi}{\sqrt{3}\mathcal{M}}\right)\right]\,.\end{split}
\end{equation}
Interestingly,  we find that the ratio between the Hubble factors at the fixed points is unchanged, i.e.  is $H_{\rm unst}/H_{\rm st}=\sqrt{2}$. In this case, the conformal transformation is $\bar{g}_{\mu\nu}=\frac{2}{\mathcal{M}^2}\frac{\partial \mathcal{L}}{\partial R}g_{\mu\nu}$ with 
\begin{equation}\chi=\frac{\alpha+\frac{3}{128\pi^2}(2\xi+1)^2}{18}\varphi+\frac{\xi-\frac{9(2\xi+1)\lambda}{64\pi^2}}{6}\phi^2\,.\end{equation}
The number of e-foldings is just as in Eq.\ \eqref{result}, and so are the slow-roll parameters as a function of N. Hence, the scalar spectral index is still $n_s\sim 1-\frac{2}{N}+O\left(\frac{1}{N^2}\right)$. These results can be readily found since, with the choice of pseudo-optimal energy scale, the quantum corrections can be seen as a redefinition of the coupling, and \eqref{result} are independent of the couplings.

\begin{figure}[ht]
  \centering 
  \includegraphics[scale=0.55]{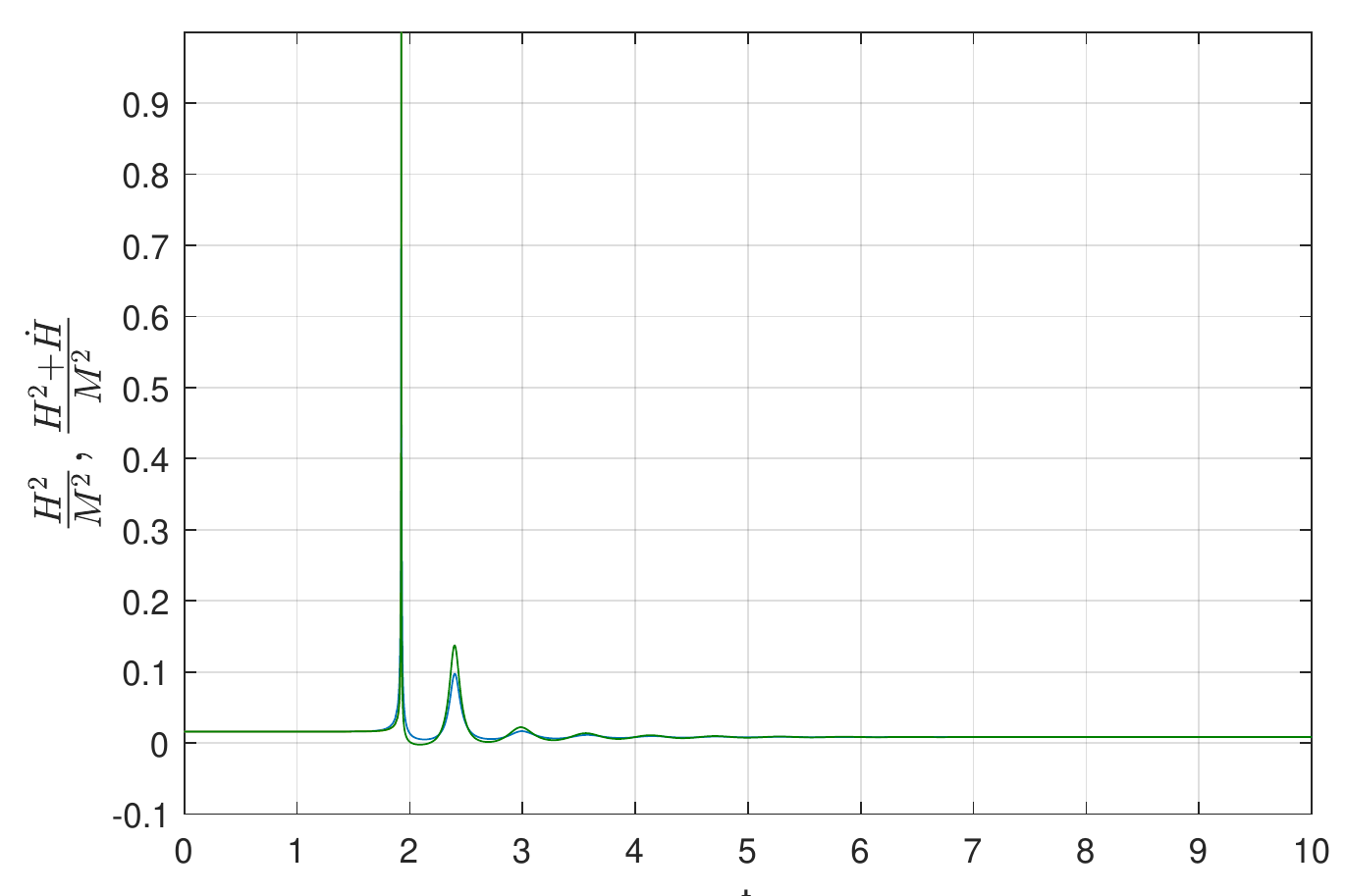} 
   \includegraphics[scale=0.5]{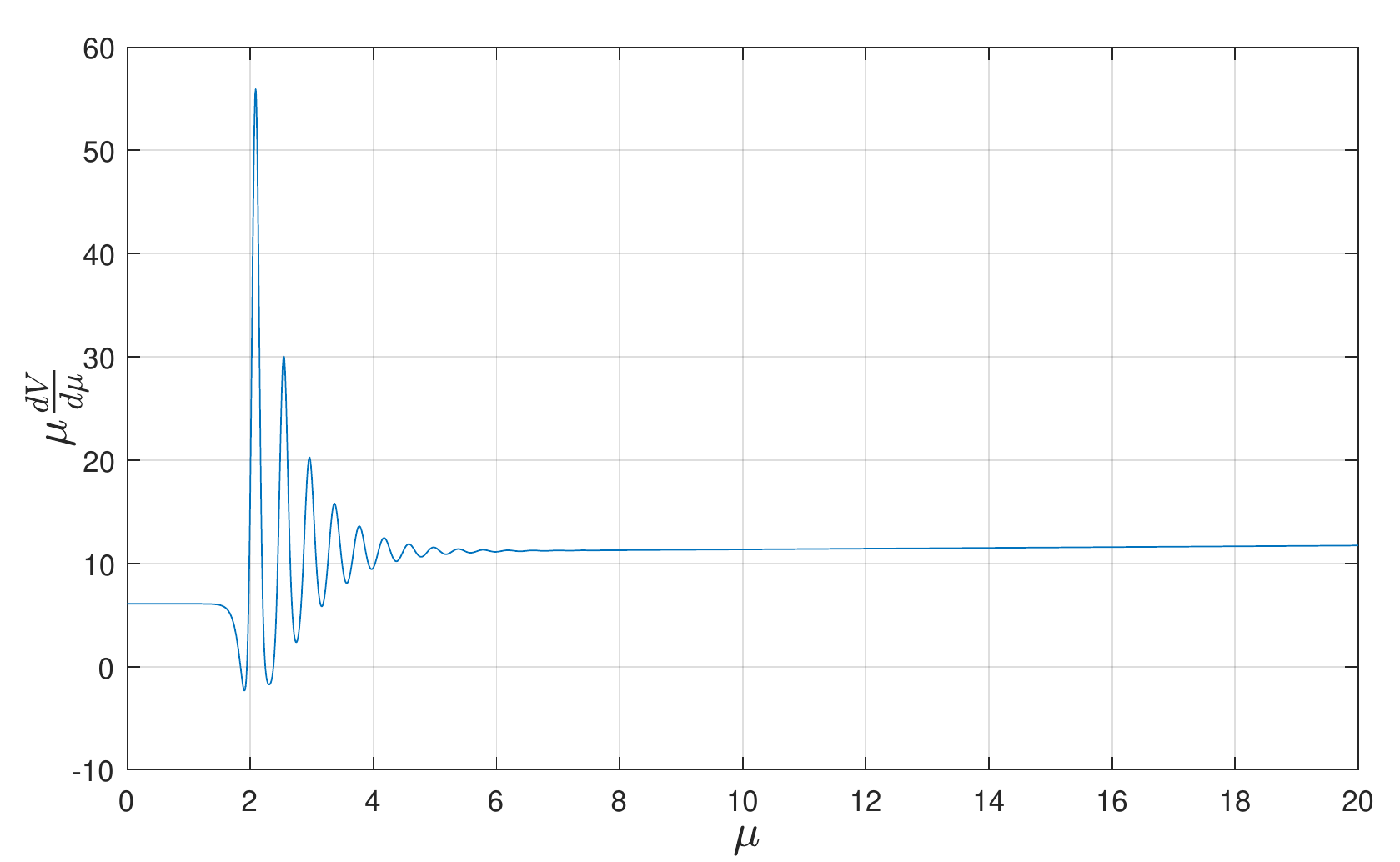} \\
   \includegraphics[scale=0.55]{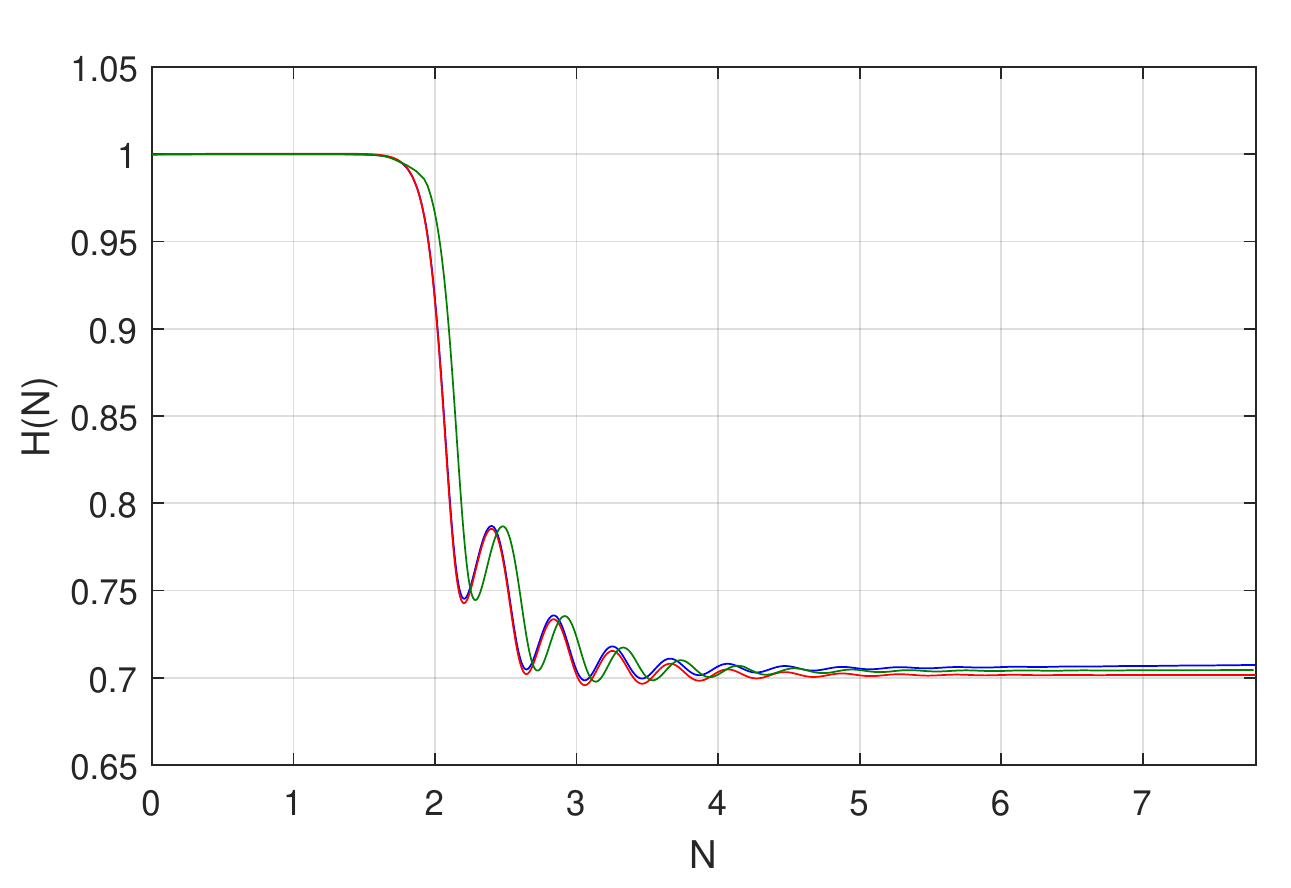}
   \includegraphics[scale=0.56]{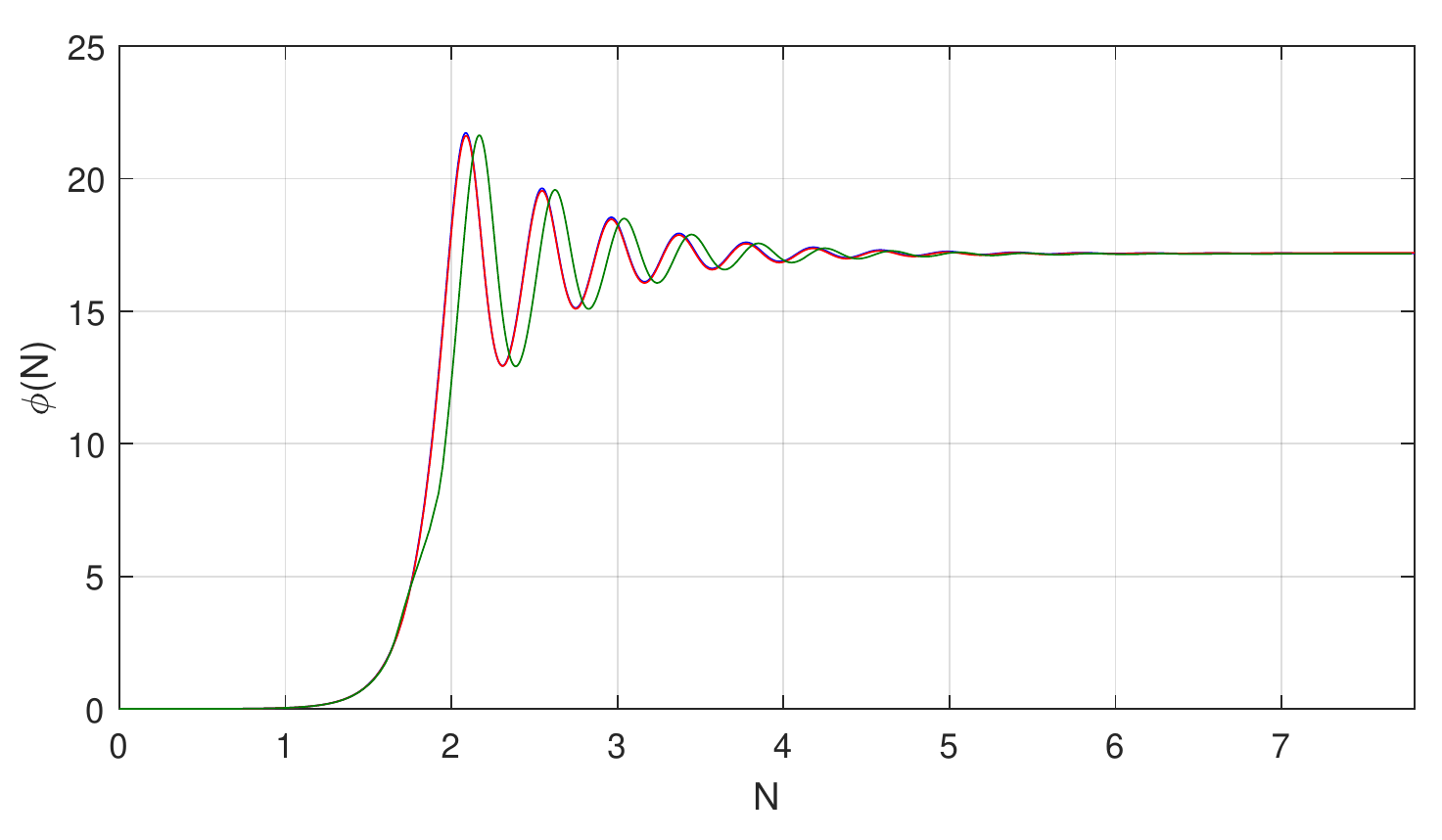}
  \caption{Top left: validity of the adiabatic approximation. The peak around $N=2$ denotes that the approximation fails. Top right:
plot of $\frac{d\Gamma}{d\mu}$ in the pseudo-optimal energy scale choice as a function of the number of e-foldings.  Bottom left: evolution of $H(N)$ from the unstable point to the stable one. The blue line is the classical evolution, the green line is the one-loop corrected one, and the red line is the one-loop corrected one implemented with the pseudo-optimal energy scale. In the classical numerical solution, couplings are chosen as $\xi=15$, $\lambda=0.1$ and $\alpha=\frac{\xi^2}{\lambda}$ and these are the initial values to solve the renormalization group equations. Bottom right:  evolution of $\phi(N)$, with the same conventions as in the bottom left plot.}
  \label{figure}
  \end{figure}

We should also quantitatively verify that the scalar spectral index matches the observations, despite the above approximation. It is possible to write the exact one-loop Lagrangian in the Jordan frame such that it is linear in $R$, as it has been done with the choice of pseudo-optimal energy scale. The problem is to write the Lagrangian with interactions between the scalar fields $\chi$ and $\phi$ (the last step in equation Eq.\ \eqref{jordan}). This is because $\chi(\varphi)$ in the exact one-loop Lagrangian is not invertible as it contains terms like $y=x\ln x$, whose inverse is the exponential of the Lambert function. Then, we only tried to find whether its contribution is numerically suppressed with some approximation: we applied the conformal transformation used with the choice of  pseudo-optimal energy scale, namely
\begin{equation}\Omega^2=\frac{2\chi}{\mathcal{M}^2}=\exp\left({\frac{\sqrt{2}\psi}{\sqrt{3}\mathcal{M}}}\right)\qquad \text{with}\qquad \chi=\frac{\alpha+\frac{3}{128\pi^2}(2\xi+1)^2}{18}\varphi+\frac{\xi-\frac{9(2\xi+1)\lambda}{64\pi^2}}{6}\phi^2\,,\end{equation} to the full one-loop potential,
and we put $R=\varphi$ and $\scalricci-\scalriem =\frac{R^2}{12}$, which is true near the unstable fixed point. The transformed Lagrangian is a reparametrization of the fields in which non-linearities in $R$ are present, but they are suppressed on-shell, thanks to the result found numerically, see Fig.\ \eqref{figure}. We find the zeroth order correction to the potential for $\frac{\phi}{\mathcal{M}}\ll1$
\begin{equation}
\label{zeroorder}
W(\phi,\psi)\sim-\frac{9\mathcal{M}^4}{4\left(\alpha+\frac{3}{128\pi^2}(2\xi+1)^2\right)}+f(\xi)\lambda^2 \mathcal{M}^3\psi\,.
\end{equation} 
The first slow-roll parameter can be computed by taking the lowest order in $\frac{\phi}{\mathcal{M}}\ll1$ of $\frac{\partial W}{\partial \psi}$. We find
\begin{equation}\epsilon=\frac{3}{4N^2}+O\left(\frac{1}{C_2^2}\right)\,,\end{equation}
with \begin{equation}C_2=\frac{\alpha+\frac{3}{128\pi^2}(2\xi+1)^2}{18}\,.\end{equation} 
This is $O(\lambda^2)$ for $\alpha=\frac{\xi^2}{\lambda}$.
The second derivative of $W$, instead, has a lowest order term proportional to $\phi^2$ so the correction will be of order $O\left(\frac{1}{N}\right)$. Thus, the second slow-roll parameter receives the correction 
\begin{equation}\eta=-\frac{1}{N}+O\left(\frac{1}{C_2N}\right)\,.\end{equation}
This is $O(\lambda)$ for $\alpha=\frac{\xi^2}{\lambda}$.
If we take a generic $\alpha$ and impose that the correction must stay within the uncertainty predicted by the Planck mission, we find the approximate lower bound for $\alpha$
\begin{equation}6\times10^{3}\lesssim\alpha\,,\end{equation}
which is five orders of magnitude less than  $\alpha=\frac{\xi^2}{\lambda}$,
 giving thus some freedom in the choice of this coupling.

\section{A note about reheating}

\noindent Reheating provides a mechanism to transfer energy from the scalar field to the Standard Model fields, which become excited and populate the Universe with all the elementary particles after the end of inflation. In our model, the backreaction of the Standard Model fields is supposed to take over the dynamical evolution of the system after it has reached the stable fixed point and then lead the Universe towards a radiation dominated era \cite{classic,tambalo}. In the simplest scenario, reheating is based on the assumption that the scalar field can decay into boson pairs $\chi$. This process can be modelled by considering the Lagrangian
\begin{equation}
\mathcal{L}=\mathcal{L}_{\rm inv}-\frac{1}{2}m_{\chi}^2 \chi^2-g^2\chi^2\phi^2-\frac{(\partial\chi)^2}{2}\,,
\end{equation}
where $\mathcal{L}_{\rm inv}$ is the scale invariant part. Expanding around the vacuum expectation value $\phi_0$ we find the relevant terms describing the decay, which take the form
\begin{equation}
\mathcal{L}_{\rm decay}\sim\frac{m_{\phi}^2}{2}\phi^2+g^2\phi_0 \phi \chi^2 \,,
\end{equation}
where $m_{\phi}^2\simeq {\lambda\phi_0^2}/{2}$ as in \cite{classic}. The decay rate is given by
\begin{equation}
\Gamma=\frac{g^2\phi_0^2}{8\pi m_{\phi}^2}=\sqrt{\frac{2}{\lambda}}\frac{g^2\phi_0}{8\pi }\,,
\end{equation}
and, in order for the field to have sufficient time to decay, we need $\Gamma \gtrsim H_0$ (where $H_0$ is the stable point value for the Hubble parameter). This provides a lower bound for the coupling $g$, which can be evaluated recalling that, at the stable fixed point, $\phi_{0}=2H_{0}\sqrt{\xi/\lambda}$, see eq.\ \eqref{effpot}. Thus, with the values inferrred in the previous sections, we have
\begin{equation}
g\gtrsim\sqrt{\frac{2\sqrt{2}\pi\lambda}{\sqrt{\xi}}}\sim 10^{-4}\,,
\end{equation}
which is hardly affected by quantum corrections since it depends only on running couplings. We also know that $\lambda$ and $\xi$ satisfy a relation at tree level (eq. 22 of \cite{classic}), which relate them to the estimated value of $H$ at the end of inflation. This can be used to  rewrite the lower bound on $g$ as
\begin{equation}
g\gtrsim 3\,\xi^{3/4} 10^{-4}\,,
\end{equation}
that shows that it has quite a strong dependence on couplings. Concerning instead upper bounds on $g$, we observe that the coupling must satisfy the perturbative conditions, dictated by the validity of the one-loop expansion, hence we expect $g$ to be much smaller than one \footnote{This, in turn, puts a constraint on $\xi$. For instance, in Higgs inflation $\xi\sim 10^{5}$, a value that is incompatible with our scenario.}. More information on the physically allowed range for $g$ could be retrieved by studying non-gaussianity, and we hope to report soon new results on this issue.

There are several alternative pictures to reheating, such as parametric resonance \cite{Kofman:1994rk}, which has been discussed for this model in refs.\ \cite{classic,tambalo}. The effects of loop corrections in this case are hard to assess without a careful analysis that goes beyond the scope of this paper but is certainly worth considering in future work.


\section{Conclusions}\label{conc}

\noindent In this paper we have studied how quantum corrections modify a classical model of inflation with spontaneous symmetry breaking of scale invariance to assess whether the viability of the model is preserved. 

In order to see the impact of quantum corrections we relied upon techniques of semi-classical gravity. This theory can be used to compute one-loop corrections in the regime in which the spacetime is slowly expanding, meaning that $\frac{\dot{a}^2(t)}{a^2(t)M^2}\ll1$, $\frac{\ddot{a}(t)}{a(t)M^2}\ll1$, where $M$ is a mass scale of the system. This  adiabatic approximation  allows to find an expansion in derivatives of the metric to the one-loop effective action. This is done up to second order and leads to a Coleman-Weinberg-like correction, where  also quadratic scalars, such as $R^2$, $R_{\alpha\beta\mu\nu}R^{\alpha\beta\mu\nu}$, $R_{\alpha\beta}R^{\alpha\beta}$, appear. The external mass scale is chosen here as field-dependent: the scaling anomaly appears via the reference value $\mu_0$ appearing in \eqref{run}. One finds also a tachyonic instability close to the onset of inflation, which cannot be handled by the heat kernel expansion, and must take into account the effect of curvature fluctuations. Some work in this case has already been done \cite{Felder:2000hj,Felder:2001kt}. 

The equations of motion are computed for generic $\mu$ and outside the tachyonic regime, including the oscillatory regime relevant to reheating. In order to solve and discuss the dynamics of the system we set $\mu^2=M^2$ in those equations which are then numerically solved along with the renormalization group differential equations. The solution has been compared to the approximated pseudo-optimal energy scale choice \cite{herr}, in which $M^2=\mu^2$ has been set readily in the Lagrangian.  This has been verified to be a good  approximation: its use allowed to simplify consistently the computation of the properties of the system and of the observables.

The main and comforting result is that there are only small deviations in the dynamics from the classical evolution. It has been verified numerically whether the adiabatic expansion holds throughout the evolution of the system: this has been proved to be true apart from a small lapse of time in which $M^2 \sim 0$ and a mild violation afterwards, so no prediction for the observables can be done in these regions of spacetime. We found that the nature of the fixed points remains unchanged: the system evolves from an unstable point to a stable one though the position of the fixed points change with respect to the classical case. The number of e-foldings has the same dependence on the fields $N\sim \ln\left(\text{const}\frac{H^2}{\phi^2}\right)$ with $\phi\sim 0$ and also the scalar spectral index remains unchanged. Quantitative deviations from the classical case are numerically suppressed so there are not consistent changes and quantum corrections do not modify the viability of the model.


\appendix

\section{Adiabatic coefficients}\label{appA}

The adiabatic coefficients  are computed according to the recursion relation 
$$\sigma(x,x')_{;\mu}a^{k;\mu}(x,x')+ka_k(x,x')=\Delta^{-1/2}(x,x')(\Delta^{1/2}(x,x')a_{k-1}(x,x'))_{;\mu}^{\,;\mu}$$
$$+\left(3\lambda(\phi^2(x')-\phi^2(x))-\frac{\xi}{3}(R(x')-R(x))-\frac{R(x')}{6}\right)a_k(x,x')\,,$$
where $\sigma$ is the geodesic interval $\frac{1}{2}(x-x')^{\alpha}(x-x')_{\alpha}$ and $\Delta(x,x')$ is the Van Vleck determinant.
These are explicitly computed in \cite{mark} by means of the heat kernel method, giving the same expression of the effective action as \eqref{regularized}.
Up to second order they are
$$a_0(x,x)=0\,,\quad a_1(x,x)=0\,,\quad a_2(x,x)=\frac{3+5\xi}{90}\Box R -\frac{1}{2}\Box\phi +\frac{\riemd{\alpha}{\beta}{\mu}{\nu}\riemu{\alpha}{\beta}{\mu}{\nu}-\ricciu{\alpha}{\beta}\riccid{\alpha}{\beta}}{180}\,.$$
The regularized effective action Eq.\ \eqref{regularized} is computed up to second order with these coefficients. Integrating by parts and truncating third and higher orders we get Eq.\eqref{onelooplagrangian} (see \cite{mark} for details).

\section{Correction to the equations of motion}\label{appB}

We report here the explicit expressions of $Q_{1}$ and $Q_{2}$:

\begin{equation}
\begin{split}
&Q_{1}=\epsilon_1\left[\frac{1}{2}\riccid{\alpha}{\beta}\ricciu{\alpha}{\beta}+2 \riemd{\rho}{0}{\gamma}{0}\ricciu{\rho}{\gamma}-\covd{0}\covd{0}R-\frac{1}{2}\Box R+\Box \riccid{0}{0}\right]+\epsilon_2 \left[\frac{1}{2}\riemd{\alpha}{\sigma}{\gamma}{\delta}\riemu{\alpha}{\sigma}{\gamma}{\delta}+\right.\\&\left.
+2R_{0}^{\,\,\rho\alpha\sigma}\riemd{0}{\rho}{\alpha}{\sigma}+4\riemd{\sigma}{0}{\gamma}{0}\ricciu{\gamma}{\sigma}-4\riccid{0}{\gamma}R^{\gamma}_{\,\,0}+4\Box \riccid{0}{0}-2\covd{0}\covd{0}R\right]
-\frac{1}{64\pi^2}\log\left(\frac{M^2}{\mu^2}\right)\times\\&\times\left[\frac{(2\xi+1)^2}{72}R^2+\frac{9\lambda^2 \phi^4}{2}+\frac{(2\xi+1)^2}{18}RR_{00}+\frac{(2\xi+1)^2}{6}HR_{,0}-\frac{\lambda(2\xi+1)}{2}R\phi^2+\right.\\&\left.-\lambda(2\xi+1)R_{00}\phi^2-\lambda(2\xi+1)6H\phi\phi_{,0}+\frac{\scalricci-\scalriem}{180}-\frac{R^{\rho}_{\,\,0}\riccid{0}{\rho}}{45}-\frac{R_{0}^{\,\,\alpha\beta\gamma}\riemd{0}{\alpha}{\beta}{\gamma}}{45}\right]+\\&-\frac{27}{4}\lambda^2\phi^4+\frac{3}{4}\lambda(2\xi+1)R\phi^2-\frac{1}{48}(2\xi+1)^2R^2-\frac{(2\xi+1)^2}{12}RR_{00}-\frac{(2\xi+1)^2}{4}HR_{,0}+\\&+\frac{3\phi^2\lambda(2\xi+1)}{2}R_{00}+9(2\xi+1)\lambda H \phi \phi_{,0}+\frac{(2\xi+1)^2}{6M^2}HRM^2_{,0}-\frac{3\lambda(2\xi+1)H}{M^2}\phi^2M^2_{,0}+\\&-\frac{2\xi+1}{6}R_{00}M^2-\frac{(2\xi+1)}{2}H M^2_{,0}+\frac{1}{90}\left[\covd{\rho}\covd{\delta}\left(\ricciu{\rho}{\delta}\log\left(\frac{M^2}{\mu^2}\right)\right)+\right.\\&\left.+2\covu{\rho}\covd{0}\left(\riccid{\rho}{0}\log\left(\frac{M^2}{\mu^2}\right)\right)-\Box\left(\riccid{0}{0}\log\left(\frac{M^2}{\mu^2}\right)\right)+4\covu{\alpha}\covu{\beta}\left(\riemd{0}{\alpha}{0}{\beta}\log\left(\frac{M^2}{\mu^2}\right)\right)+\right.\\&\left.-\frac{(2\xi+1)R_{00}}{6M^2}(\scalricci-\scalriem)+\frac{(2\xi+1)H}{2M^4}M^2_{,0}\left(\scalricci-\scalriem\right)+\right.\\&\left.-\frac{(2\xi+1)H}{2M^2}\left(\scalricci-\scalriem\right)_{,0}\right]=0\,,
\end{split}
\end{equation}

\begin{equation}
\begin{split}
&Q_{2}=
\frac{1}{64\pi^2}\left[\left(36\lambda^2 \phi^3-2\lambda(2\xi+1)R\phi\right)+\left(\log\left(\frac{M^2}{\mu^2}\right)-\frac{3}{2}\right)+6\lambda \phi M^2+\right.\\&\left.-\frac{1}{90}(\scalricci-\scalriem)\frac{6\lambda \phi}{M^2}\right].
\end{split}
\end{equation}

\section{Solution of the renormalization group equations}\label{appC}

Eqs.\ \eqref{runcoupling} can be easily integrated, and we find
\begin{eqnarray}
\label{run}
 \epsilon_{1,2}(\mu)&=&\epsilon_{1/2,0}\pm \frac{\ln (\mu/\mu_0)}{2880\pi^2}\,,\\\nonumber
\lambda(\mu)&=&\frac{\lambda_0}{1-\frac{9}{8\pi^2}\ln\left(\frac{\mu}{\mu_0}\right)}\,,\\\nonumber
2\xi(\mu)+1&=&(2\xi_0+1)\left(1-\frac{9}{8\pi^2}\ln\left(\frac{\mu}{\mu_0}\right)\right)^{-\lambda_0/3}\,,\\\nonumber 
\alpha(\mu)&=&\alpha_0-\frac{\pi^2(2\xi_0+1)^2}{36(1-2\lambda_0/3)}+\frac{\pi^2(2\xi_0+1)^2}{36(1-2\lambda_0/3)}\left(1-\frac{9}{8\pi^2}\ln\left(\frac{\mu}{\mu_0}\right)\right)^{-2\lambda_0/3+1} \,,
\end{eqnarray}
where the solution for $\alpha(\mu)$ and $\xi(\mu)$ is valid for $\mu\ll\mu_0 e^{8\pi^2/9}$. This result is discussed in the main text.



\begin{thebibliography}{85}
 			\bibitem{planck} P.~A.~R.~Ade {\it et al.} [Planck Collaboration],
  Astron.\ Astrophys.\  {\bf 594} (2016) A13.
  
 		\bibitem{star}A. A. Starobinsky, 
		Proceedings of the 2nd Seminar on Quantum Gravity, Moscow, 13 15 October 1981, pp. 5872, (INR Press, Moscow,
 		1982). Reprinted in: Markov, M.A. and West, P.C., eds., Quantum Gravity, (Plenum Press,	New York, 1984), pp. 103128;
		
 		A. A. Starobinsky, Phys. Lett. B, 91, 99102, (1980).
 		\bibitem{defelice} A.\ De Felice, S.\ Tsujikawa, Living Rev. Relativity, 13, (2010).
		
 		\bibitem{classic} M.~Rinaldi and L.~Vanzo,
  Phys.\ Rev.\ D {\bf 94} (2016) no.2,  024009.
  
  \bibitem{tambalo}
   G.~Tambalo and M.~Rinaldi,
  Gen.\ Rel.\ Grav.\  {\bf 49} (2017) no.4,  52.
  
  		\bibitem{cooper} F.~Cooper and G.~Venturi,
  Phys.\ Rev.\ D {\bf 24} (1981) 3338.
  
  \bibitem{Gundhi:2018wyz}
  A.~Gundhi and C.~F.~Steinwachs,
  arXiv:1810.10546 [hep-th].
  
  \bibitem{Kurkov:2013gma} 
  M.~A.~Kurkov and M.~Sakellariadou,
  JCAP {\bf 1401}, 035 (2014).
  
  \bibitem{Myrzakulov:2016tsz} 
  R.~Myrzakulov, S.~Odintsov and L.~Sebastiani,
  Nucl.\ Phys.\ B {\bf 907}, 646 (2016).
 
   
  \bibitem{inf3} M.~Rinaldi, L.~Vanzo, S.~Zerbini and G.~Venturi,
  Phys.\ Rev.\ D {\bf 93} (2016) 024040.
 
 \bibitem{inf1}  M.~Rinaldi, G.~Cognola, L.~Vanzo and S.~Zerbini,
  Phys.\ Rev.\ D {\bf 91} (2015) no.12,  123527.

\bibitem{inf2}  M.~Rinaldi, G.~Cognola, L.~Vanzo and S.~Zerbini,
  JCAP {\bf 1408} (2014) 015.

 \bibitem{lalak} Z.~Lalak and L.~Nakonieczny,
  Phys.\ Dark Univ.\  {\bf 15} (2017) 125.


\bibitem{bh2}  G.~Cognola, M.~Rinaldi, L.~Vanzo and S.~Zerbini,
  Phys.\ Rev.\ D {\bf 91} (2015) 104004.

\bibitem{bh1}  G.~Cognola, M.~Rinaldi and L.~Vanzo,
  Entropy {\bf 17} (2015) 5145.
%
 \bibitem{shapiro} ``Effective action in quantum gravity'', I.\ L.\ Buchbinder, S.\ D.\ Odintsov, I.\ L.\ Shapiro, IOP Publishing Ltd, 1992.
 
 \bibitem{parker} ``Quantum Field Theory in Curved Spacetime: Quantized Fields and Gravity'', L.\ Parker, D.\ Toms, Cambridge University Press (2009).


\bibitem{Kubo:2018kho} 
  J.~Kubo, M.~Lindner, K.~Schmitz and M.~Yamada,
  arXiv:1811.05950 [hep-ph].
  
  			
 \bibitem{scaleinv2}
 		P.~G.~Ferreira, C.~T.~Hill and G.~G.~Ross,
 		Phys.\ Rev.\ D {\bf 98} (2018) no.11,  116012.
 \bibitem{scaleinv3}Mikhail Shaposhnikov and Daniel Zenhausern. Phys. Lett.,
 		B671:162-166, 2009.
 \bibitem{scaleinv1}D. M. Ghilencea, Phys. Rev. D 93, no. 10, 105006 (2016)
 		; D.~M.~Ghilencea, Z.~Lalak and P.~Olszewski,
  Eur.\ Phys.\ J.\ C {\bf 76} (2016) no.12,  656.
 \bibitem{birrell}``Quantum Fields in Curved Space``,
 		N.\ D.\ Birrell, P.\ C.\ W.\ Davies, Cambridge University Press (1984).
 
 \bibitem{WB95}
 W.~A.~Bardeen, ``On Naturalness in the Standard Model'', Presented at the 1995 Ontake Summer Institute, Ontake Mountain, Japan 1995, FERMILAB-CONF-95-391-T
 
\bibitem{DeWitt:2003pm} 
  B.~S.~DeWitt,
  Int.\ Ser.\ Monogr.\ Phys.\  {\bf 114}, 1 (2003).

 \bibitem{Felder:2000hj} 
  G.~N.~Felder, J.~Garcia-Bellido, P.~B.~Greene, L.~Kofman, A.~D.~Linde and I.~Tkachev,
  Phys.\ Rev.\ Lett.\  {\bf 87}, 011601 (2001).
 
\bibitem{Felder:2001kt}
  G.~N.~Felder, L.~Kofman and A.~D.~Linde,
  Phys.\ Rev.\ D {\bf 64} (2001) 123517.
  \bibitem{Ruf:2017bqx}
  M.~S.~Ruf and C.~F.~Steinwachs,
  Phys.\ Rev.\ D {\bf 97} (2018) no.4,  044049.
 \bibitem{Bamba:2014mua} 
  K.~Bamba, G.~Cognola, S.~D.~Odintsov and S.~Zerbini,
  Phys.\ Rev.\ D {\bf 90}, no. 2, 023525 (2014).
  				 
 \bibitem{herr} M.~Herranen, A.~Hohenegger, A.~Osland and A.~Tranberg,
 				Phys.\ Rev.\ D {\bf 95} (2017) no.2,  023525. 
 
 
 \bibitem{guth}Alan H. Guth, Phys. Rev. D 23, 347 (1981).
 
 \bibitem{Karam:2018mft} A.~Karam, T.~Pappas and K.~Tamvakis ``Nonminimal Coleman--Weinberg Inflation with an $R^2$ term,'' arXiv:1810.12884 [gr-qc] (to appear in JCAP).
 \bibitem{dabrovski}
 				 M.~P.~Dabrowski, J.~Garecki and D.~B.~Blaschke,
 				Annalen Phys.\  {\bf 18} (2009) 13.
 				\bibitem{herzberg}
 				 M.~P.~Hertzberg,
 				Phys.\ Lett.\ B {\bf 745} (2015) 118.
 				\bibitem{mark} T.~Markkanen and A.~Tranberg,
 				JCAP {\bf 1211} (2012) 027.
 				
 			\bibitem{quantum1} A.\  Kamenshchik and C.\ Steinwachs,
 			Phys.\ Rev.\ D {\bf 91} (2015) no.8,  084033.
 			
 			\bibitem{Ruf:2017xon}
 			M.~S.~Ruf and C.~F.~Steinwachs,
 			Phys.\ Rev.\ D {\bf 97} (2018) no.4,  044050.
 					\bibitem{quantum2}  N.~Ohta,
 					PTEP {\bf 2018} (2018) no.3,  033B02.
					
					\bibitem{maxeq} M.~Rinaldi,
  Eur.\ Phys.\ J.\ Plus {\bf 133} (2018) no.10,  408.
					

\bibitem{canko} D.~D.~Canko, I.~D.~Gialamas and G.~P.~Kodaxis,
  ``A simple $F({\cal R},\phi)$ deformation of Starobinsky inflationary model,''
  arXiv:1901.06296 [hep-th].

\bibitem{Kofman:1994rk}
L.~Kofman, A.~D.~Linde and A.~A.~Starobinsky,
Phys.\ Rev.\ Lett.\  {\bf 73} (1994) 3195
[hep-th/9405187].

\end{thebibliography}
\end{document}